\documentclass[twocolumn,preprintnumbers]{revtex4}
\usepackage{graphicx}
\usepackage{dcolumn}
\usepackage{bm}
\usepackage{subfigure}
\usepackage{float}
\begin{document}
\newcommand{\ds}{\displaystyle}
\newcommand{\be}{\begin{equation}}
\newcommand{\en}{\end{equation}}
\newcommand{\bea}{\begin{eqnarray}}
\newcommand{\ena}{\end{eqnarray}}
\title{$(N+1)$-dimensional Lorentzian evolving wormholes supported by polytropic matter}
\author{Mauricio Cataldo}
\altaffiliation{mcataldo@ubiobio.cl} \affiliation{Departamento de
F\'\i sica, Facultad de Ciencias, Universidad del B\'\i o-B\'\i o,
Avenida Collao 1202, Casilla 5-C, Concepci\'on, Chile.}
\author{Fernanda Ar\'{o}stica}
\altaffiliation{ferarostica@udec.cl} \affiliation{Departamento de F\'{\i}sica, Universidad de Concepci\'{o}n,\\
Casilla 160-C, Concepci\'{o}n, Chile.}
\author{Sebastian Bahamonde}
\altaffiliation{sbahamonde@udec.cl} \affiliation{Departamento de F\'{\i}sica, Universidad de Concepci\'{o}n,\\
Casilla 160-C, Concepci\'{o}n, Chile.}\date{\today}
\smallskip
\maketitle 

{\bf Abstract}: In this paper we study $(N+1)$-dimensional evolving
wormholes supported by energy satisfying a polytropic equation of
state. The considered evolving wormhole models are described by a
constant redshift function and generalizes the standard flat
Friedmann-Robertson-Walker spacetime. The polytropic equation of
state allows us to consider in $(3+1)$-dimensions generalizations of
the phantom energy and the generalized Chaplygin gas sources.

\section{Introduction}
Wormholes are exotic solutions of Einstein's field equations
representing tunnels through space-time connecting two different
regions of our Universe or even another Universe.  The focus of
studying wormhole geometries has intensively increased since the
publication of works of Morris and Thorne~\cite{1,1A,Visser}, where
it was proposed the possibility of the existence of traversable
wormholes allowing to travel through space and time. The idea of
such a travel always has been interesting for the humanity so a lot
of works has studied wormhole models with different types of matter
supporting
them~\cite{StaticWH,StaticWHA,StaticWHB,StaticWHC,StaticWHD,StaticWHE,StaticWHF,StaticWHG,StaticWHH,StaticWHI,StaticWHJ,StaticWHK,StaticWHL,StaticWHM,StaticWHN,StaticWHO},
or in alternative theories of
gravity~\cite{Alternative,AlternativeA,AlternativeB,AlternativeC}.
Most considered models are stationary spherically symmetric
wormholes. The matter responsible for sustaining a traversable
wormhole is exotic since it violates the standard null and weak
energy conditions. In other words, it would only be possible to
cross such a stationary wormhole if exotic matter with negative
energy density sustain it. Constructed static spherically symmetric
wormholes are described by the metric~\cite{1,Visser}
\begin{eqnarray}\label{metrica1}
ds^{2}&=&-e^{\Phi(r)}dt^{2}+\frac{dr^{2}}{1-\frac{b(r)}{r}}+d\Omega^{2},
\end{eqnarray}
where $\Phi(r)$ is the redshift function and $b(r)$ the shape
function. These functions depend on the form assumed for the
energy-momentum tensor coupled to the metric ~(\ref{metrica1}). In
order to have a wormhole geometry the redshift and shape functions
must obey some specific conditions such as for example $\Phi(r)$
must be finite everywhere for guaranteeing the absence of horizons
and singularities in the space-time, $b(r=r_{0})=r_{0}$ at a throat,
$b(r)/r\leq 1$ and, in order to have an asymptotically Minkowskian
space-time, the condition $b(r)/r\rightarrow 0$ at $r\rightarrow
\infty$  must be required.

This wormhole concept can be extended to time-dependent wormhole
geometries. The metric considered for describing such an evolving
wormhole may be written in the form
\begin{eqnarray}\label{15}
ds^{2}&=&-e^{-2\Phi(t,r)}dt^{2}+a^{2}(t)\left[\frac{dr^{2}}{1-\frac{b(r)}{r}}+
r^{2}d\Omega^{2} \right],
\end{eqnarray}
where the redshift function now depends on the time and radial
coordinates, and the new function $a(t)$ is the scale factor. This
function controls the expansion of the wormhole, and its evolution
is dictated by Einstein's field equations.

In this paper we want to discuss wormhole models supported by
polytropic phantom energy, resulting in an extension of the
polytropic wormholes discussed in Ref.~\cite{jamil}. The
generalization goes in two ways. First, we are interested in
obtaining a static $(N+1)$-dimensional extension of zero-tidal-force
wormhole models studied in Ref.~\cite{jamil}. The second extension
consists in their dynamic generalization by introducing a scale
factor with the help of the metric~(\ref{15}).

It must be noticed, for example, that in four dimensional
spherically symmetric spacetimes analytical solutions for polytropic
star configurations are known just for a few particular values of
the polytropic index (excluding the linear equation of state
$p=\omega \rho$). An interesting discussion involving polytropic
equations of state for general relativistic stars is given in
Ref.~\cite{Nilsson}. In Ref.~\cite{4}, authors found solutions for a
compressible polytropic fluid sphere in gravitational equilibrium. A
study on polytropic stars in three-dimensional space-time is made
in~\cite{Sa}. It therefore seems of interest to find spherically
symmetric gravitational models supported with matter fields obeying
a polytropic equation of state.

The organization of the paper is as follows: In Sec. II we present
the dynamical field equations for wormhole models with a matter
source obeying a polytropic equation of state. In Sec. III solutions
to these field equations are studied, and in Sec. IV we conclude
with some remarks.

\section{$(N+1)$-dimensional evolving wormholes}
Let us now consider the $(N+1)$-dimensional extension of the
metric~(\ref{15}), with a vanishing redshift function, described in
comoving coordinates $(t,r, \theta_1,...\theta_{N-1})$ by the metric
\begin{eqnarray}\label{2}
ds^{2}&=&-dt^{2}+a^{2}(t)\left[\frac{dr^{2}}{1-\frac{b(r)}{r}}+r^{2}d\Omega^{2}_{N-1}
\right], \,\,\, \,\,\,\,
\end{eqnarray}
where $d\Omega^{2}_{N-1}$ is the metric on a $(N-1)$ sphere. Since
$\Phi(t,r)=0$ this metric describes a zero-tidal-force wormholes,
ensuring the absence of horizons and singularities in the considered
space-time. It is clear that the metric~(\ref{2}) becomes an
$(N+1)$-dimensional static zero-tidal-force wormhole if
$a(t)\rightarrow \textrm{constant}$ and, as $b(r)\rightarrow 0$, it
becomes a flat $(N+1)$-dimensional extension of the
Friedmann-Robertson-Walker (FRW) metric.

In order to simplify the analysis we shall rewrite the metric as
\begin{eqnarray}\label{metric}
ds^{2}&=&-\theta^{(t)}\theta^{(t)}+\theta^{(r)}\theta^{(r)}+\sum_{i=1}^{N-1}{\theta^{(\theta_{i})}\theta^{(\theta_{i})}},
\end{eqnarray}
where the following $(N+1)$-dimensional proper orthonormal basis of
one-forms
\begin{eqnarray}
\theta^{(t)}&=& dt,\label{basis}\\
\theta^{(r)}&=&a(t)\frac{dr}{\sqrt{1-\frac{b(r)}{r}}},\\
\theta^{(\theta_{1})}&=&a(t)r d_{\theta_{1}},\\
\theta^{(\theta_{2})}&=&a(t)r \sin{\theta_{1}}d_{\theta_{2}},\\
\theta^{(\theta_{N-1})}&=&a(t)r
\prod_{i=1}^{N-2}{\sin{\theta_{i}}d\theta_{N-1}}, \label{basis'}
\end{eqnarray}
has been introduced. We shall consider a matter source described by
an inhomogeneous and anisotropic fluid with a diagonal
energy-momentum tensor. Then the only nonzero components of the
energy-momentum tensor in the basis (\ref{basis})-(\ref{basis'})
are:
\begin{eqnarray}\label{EMT1}
T_{(t)(t)}&=&\rho(t,r),\\
T_{(r)(r)}&=&p_{r}(t,r),\\
T_{(\theta)(\theta)}&=...=&T_{(\theta_{N-1})(\theta_{N-1})}=p_{l}(t,r),
\label{EMT2}
\end{eqnarray}
where $p_{r}(t,r)$ and $p_{l}(t,r)$  are the radial and lateral
pressures and $\rho(t,r)$ the energy density of the fluid for an
observer who always remain at rest at constant $r$, $\theta_1$, ...,
$\theta_{N-1}$.

Thus for the metric~(\ref{2}), and the energy-momentum tensor given
by expressions~(\ref{EMT1})-(\ref{EMT2}), the Einstein's equation
with cosmological constant $\Lambda$ may be written in the following
form:
\begin{eqnarray}
\frac{(N-1)}{2a^{2}r^{3}}[rb'(r)+(N-3)b(r)]+&&\nonumber\\
\frac{N(N-1)}{2}H^{2}&=&\kappa \rho+\Lambda,  \label{N1} \\
-\frac{(N-1)(N-2)}{2}H^{2}-(N-1)\frac{\ddot{a}}{a}-&&\nonumber \\
(N-1)(N-2)\frac{b}{2r^{3}a^{2}}&=&\kappa p_{r}-\Lambda, \label{N2} \\
-\frac{(N-1)(N-2)}{2}H^{2}-(N-1)\frac{\ddot{a}}{a}-&&\nonumber \\
\frac{(N-2)}{2r^{3}a^{2}}[rb'(r)+(N-4)b(r)]&=&\kappa p_{l}-\Lambda, \label{N3}
\end{eqnarray}
where $\kappa=8\pi G$, $H=\dot{a}(t)/a(t)$ and a prime and an
overdot denote differentiation $d/dr$ and $d/dt$ respectively.

Using the conservation of the energy-momentum tensor $T^{\mu}_{\
\nu;\mu}=0$, we obtain the following equations:
\begin{eqnarray}
\frac{\partial \rho}{\partial t} + H\left[p_{r}+(N-1)p_{l}+N\rho\right]&=&0, \label{cons1}\\
\frac{\partial p_{r}}{\partial r}-\frac{(N-1)}{r}(p_{l}-p_{r})&=&0.\label{cons2}
\end{eqnarray}
Additionally, in order to close the system~(\ref{N1})-(\ref{N3}) a
constitutive equation, relating the radial pressure $p_r(t,r)$ to
the energy density $\rho(t,r)$, can be introduced. We are interested
in studying polytropic equations of state. Strictly speaking, this
means that one should consider the equations of state
$p_{r}(t,r)=\omega\rho(t,r)^{\gamma}$, where $\omega$ and  $\gamma$
are the state parameter and the polytropic index of the cosmic fluid
respectively. For $\gamma=1$ we obtain the standard barotropic
equation of state.

In this paper we are interested in considering the energy density as
a function with separated variables, i.e. having the form
\begin{eqnarray}\label{rhosep}
\rho(t,r)&=&\rho_{c}(t) \rho_{w}(r).
\end{eqnarray}
where we have introduced the functions $\rho_{c}$ and $\rho_{w}$,
which depend on the time and radial coordinates respectively. Thus
the radial pressure should take the form
\begin{eqnarray}\label{prsep}
p_{r}(t,r)= \omega\rho_{c}^\gamma(t)\rho_{w}^{\gamma}(r).
\end{eqnarray}
Unfortunately, it can be shown that in general such polytropic
solutions do not exist. Effectively, by putting Eqs.~(\ref{rhosep})
and~(\ref{prsep}) into Eqs.~(\ref{N1}) and~(\ref{N2}) we conclude,
after a little algebra, that for having evolving wormholes the
condition $\gamma=1$ must be fulfilled, while for solutions with
$\gamma \neq 1$ the condition $a(t)=const$ must be required.
Barotropic evolving wormhole geometries for $\gamma =1$ were
considered in Ref.~\cite{cataldo}, and polytropic static wormholes
with $a(t)=const$ were studied in Ref.~\cite{jamil}.

In the following, we shall study polytropic solutions for the
considered field equations by introducing a special equation of
state, which we shall dub ``partially polytropic equation of state".
In order to do this we shall consider that the radial pressure can
be written as
\begin{eqnarray}
p_{r}(t,r)&=& \omega\rho_{c}(t)\rho_{w}^{\gamma}(r),\label{pr}
\end{eqnarray}

Clearly, if $\rho_{c}(t)=const$ we obtain the polytropic equation of
state $p_r(r)=\omega \rho(r)^{\gamma}$ considered in
Ref.~\cite{jamil}. The constants $\omega$ and $\gamma$ are the
barotropic state parameter and the polytropic index of the proposed
cosmic fluid respectively. Notice that from Eqs.~(\ref{cons2}),
(\ref{rhosep}) and~(\ref{pr}) we conclude that the lateral pressure
may be written as
\begin{eqnarray}\label{plsep}
p_{l}(t,r)&=&p_{lc}(t)p_{lw}(r).
\end{eqnarray}
By introducing Eqs.~(\ref{rhosep})-(\ref{plsep}) into the field
Eqs.~(\ref{N1})-(\ref{N3}) we can rewrite Einstein's equations in
the following form:
 \begin{eqnarray}
\frac{(N-1)}{2a^{2}r^{3}}[rb'(r)+(N-3)b(r)]+&&\nonumber\\
\frac{N(N-1)}{2}H^{2}&=&\kappa \rho_{w}(r)\rho_{c}(t)+\nonumber\\
&&\Lambda,  \label{N1'} \\
-\frac{(N-1)(N-2)}{2}H^{2}-(N-1)\frac{\ddot{a}}{a}-&&\nonumber \\
(N-1)(N-2)\frac{b}{2r^{3}a^{2}}&=&\kappa \omega\rho_{c}(t)\rho_{w}^{\gamma}(r)-\nonumber\\
&&\Lambda, \label{N2'} \\
-\frac{(N-1)(N-2)}{2}H^{2}-(N-1)\frac{\ddot{a}}{a}-&&\nonumber \\
\frac{(N-2)}{2r^{3}a^{2}}[rb'(r)+(N-4)b(r)]&=&\kappa p_{lc}(t)p_{lw}(r)-\nonumber\\
&&\Lambda. \label{N3'}
\end{eqnarray}
\section{Wormhole solutions}
For solving the field equations we shall use the method of separable
variables. By dividing Eqs.~(\ref{N1'})-(\ref{N2'}) by
$\rho_{c}(t)$, and Eq.~(\ref{N3'}) by $p_{lc}(t)$ we obtain
\begin{eqnarray}
\Big[\frac{N(N-1)}{2}H^{2}-\Lambda\Big]\rho^{-1}_{c}(t)&=&\nonumber\\
\kappa \rho_{w}(r)-\frac{(N-1)}{2a^{2}(t)\rho_{c}(t)r^{3}}[rb'(r)+(N-3)b(r)],\\
\Big[\Lambda-\frac{(N-1)(N-2)}{2}H^{2}-(N-1)\frac{\ddot{a}}{a}\Big]\rho^{-1}_{c}(t)&=&\nonumber\\
\ \ \ \ \ \ \ \ \ \ \ \ \ \ \frac{b(r)(N-1)(N-2)}{2r^{3}\rho_{c}(t)a^{2}(t)}+\kappa \omega\rho_{w}^{\gamma}(r),\\
\Big[\Lambda-\frac{(N-1)(N-2)}{2}H^{2}-(N-1)\frac{\ddot{a}}{a}\Big]p^{-1}_{lc}(t)&=&\nonumber\\
\frac{(N-2)(rb'(r)+(N-4)b(r))}{2r^{3}p_{lc}(t)a^{2}(t)}+\kappa p_{lw}(r).
\end{eqnarray}
The LHS of the above equations depends only on $t$, so by
introducing the conditions
\begin{eqnarray}
\rho_{c}(t)&=&\tilde{C}_{1}a^{-2}(t),\label{rhoc}\\
p_{lc}(t)&=&\tilde{C}_{2}a^{-2}(t),\label{plc}
\end{eqnarray}
where $\tilde{C}_{1}$ and $\tilde{C}_{2}$ are constants, the RHS of
these equations depends only on the $r$-coordinate. This allows us
to rewrite the field equations in the form
\begin{eqnarray}
\Big[\frac{N(N-1)}{2}H^{2}-\Lambda\Big]a^{2}(t)&=&\nonumber\\
\kappa\tilde{C}_{1} \rho_{w}(r)-\frac{(N-1)}{2r^{3}}[rb'(r)+(N-3)b(r)]=-3C_{3},\label{sep1}\\
\Big[\Lambda-\frac{(N-1)(N-2)}{2}H^{2}-(N-1)\frac{\ddot{a}}{a}\Big]a^{2}(t)&=&\nonumber\\
 \frac{b(r)(N-1)(N-2)}{2r^{3}}+\kappa \omega\tilde{C}_{1}\rho_{w}^{\gamma}(r)=-Q,\label{sep2}\\
\Big[\Lambda-\frac{(N-1)(N-2)}{2}H^{2}-(N-1)\frac{\ddot{a}}{a}\Big]a^{2}(t)&=&\nonumber\\
\frac{(N-2)(rb'(r)+(N-4)b(r))}{2r^{3}}+\kappa \tilde{C}_{2}
p_{lw}(r)=-Q,\label{sep3}
\end{eqnarray}
where we have introduced the separation constants $C_{3}$ and $Q$,
which are independent of the coordinates $t$ and $r$.

Let us first solve the time dependent part of these equations. From
Eq.~(\ref{sep1}) we obtain that the scale factor of the universe is
given by \small
\begin{eqnarray}\label{scalefactor}
a(t)=a_{0}\left(3NC_{3}(N-1)e^{\mp\sqrt{\frac{2\Lambda
t}{N(N-1)}}}+e^{\pm\sqrt{\frac{2\Lambda t}{N(N-1)}}}\right), \, \,
\end{eqnarray}

\normalsize for $\Lambda \neq 0$, and takes the form
\begin{eqnarray}\label{scalefactor const}
a(t)=C_1 \pm \sqrt{-\frac{C_3}{N(N-1)}} \,  t
\end{eqnarray}
for $\Lambda=0$, where $a_{0}$ and $C_1$ are integration constants.
It becomes clear that we have accelerated (decelerated) expansion
for $\Lambda \neq 0$, while for $\Lambda =0$ and $C_3 <0$ the space
expands with constant velocity.

By putting Eq.~(\ref{scalefactor}) into Eq.~(\ref{sep2}) (or into
Eq.~(\ref{sep3})) we obtain the following relationship for the
separation constants:
\begin{eqnarray}\label{condition}
Q&=&\frac{3(2-N)C_{3}}{N}.
\end{eqnarray}
Now let us consider the static part of
Eqs.~(\ref{sep1})-(\ref{sep3}). By replacing Eqs.~(\ref{sep2})
and~(\ref{condition}) into Eq.~(\ref{sep1}) we obtain the
differential equation \footnotesize
\begin{eqnarray}
(\kappa\tilde{C}_{1})^{\frac{(\gamma-1)}{\gamma}}[-\omega]^{-1/\gamma}\Big[\frac{(N-1)(N-2)b(r)}{2r^{3}}
-\frac{3(N-2)C_{3}}{N}\Big]^{1/\gamma}&&\nonumber\\
-\frac{(N-1)}{2r^{3}}\left[rb'(r)+(N-3)b(r)\right]=-3C_{3}.\,\,
&&
\end{eqnarray}
\normalsize The above equation does not have analytical solutions
for an arbitrary $\gamma$. In order to find analytical solutions for
any polytropic index $\gamma$ we shall make $C_{3}=0$, which implies
from Eq. (\ref{condition}) that
$Q=0$, obtaining 
\begin{eqnarray}\label{b(r)}
b(r)=\frac{2r^{3-N}}{N-1} \times
\,\,\,\,\,\,\,\,\,\,\,\,\,\,\,\,\,\,\,\,\,\,\,\,\,\,\,\,\,\,
\nonumber
\\ \left[F+\frac{(N-2)^{1/\gamma}}{N}\frac{(\kappa \tilde{C}_{1})
^{\frac{(\gamma-1)}{\gamma}}}{(-\omega)^{1/\gamma}}r^{\frac{N(\gamma-1)}{\gamma}}\right]^{\frac{\gamma}{(\gamma-1)}},
\,\,
\end{eqnarray}
where $F$ is an integration constant. For $N=3$ we obtain the
solution discussed in Ref. \cite{jamil}.

Now by replacing Eq.~(\ref{b(r)}) into Eq.~(\ref{sep1}), and by
using Eqs.~(\ref{rhosep}) and~(\ref{rhoc}), we obtain for the energy
density
\begin{eqnarray}\label{rho(r)}
\rho(t,r)&=&\tilde{C}_{1}\left(\frac{(N-2)r^{-N}}{(-\omega)\kappa \tilde{C}_{1}}\right)^{1/\gamma}\Big[F+\nonumber\\
&&\frac{(N-2)^{1/\gamma}}{N}\frac{(\kappa\tilde{C}_{1})
^{\frac{(\gamma-1)}{\gamma}}}{(-\omega)^{1/\gamma}}r^{\frac{N(\gamma-1)}{\gamma}}\Big]^{\frac{1}{(\gamma-1)}}a^{-2}(t).
\, \, \, \, \, \,  \, \, \, \, \, \, \,
\end{eqnarray}
By taking into account Eq.~(\ref{pr}) we have that the radial
pressure is given by
\begin{eqnarray}\label{prtotal}
p_{r}(t,r)=\frac{(2-N)r^{-N}}{\kappa} \times \,\,\,\,\,\,\,\,\,\,\,\,\,\,\,\,\,\,\,\,\,\,\,\,\,\,\,\, \nonumber\\
\Big[F+\frac{(N-2)^{1/\gamma}}{N}\frac{\kappa
^{\frac{(\gamma-1)}{\gamma}}}{(-\omega)^{1/\gamma}}r^{\frac{N(\gamma-1)}{\gamma}}\Big]^{\frac{\gamma}{(\gamma-1)}}a^{-2}(t),
\,\,\,\,\,\,\,\,
\end{eqnarray}
and from Eqs.~(\ref{plsep}), (\ref{plc}), (\ref{sep3})
and~(\ref{b(r)}),
we obtain for the lateral pressure 
\begin{eqnarray}\label{pl}
p_{l}(t,r)=\frac{(N-2)r^{-N}}{N-1} \times
\nonumber \\
\left(F+\frac{(N-2)^{1/\gamma}}{N}\frac{(\kappa\tilde{C}_{1})
^{\frac{(\gamma-1)}{\gamma}}}{(-\omega)^{1/\gamma}}r^{\frac{N(\gamma-1)}{\gamma}}\right)^{\frac{1}{(\gamma-1)}}
\times \nonumber \\
\Big[F+\frac{(1-N)(N-2)^{1/\gamma}}{N}\frac{(\kappa\tilde{C}_{1})^{\frac{(\gamma-1)}{\gamma}}}{(-\omega)^{1/\gamma}}r^{\frac{N(\gamma-1)}{\gamma}}\Big]a^{-2}(t).
\end{eqnarray}
In this case the constraint $C_{3}=Q=0$ implies that the scale
factor (\ref{scalefactor}) becomes
\begin{eqnarray}\label{scalef}
a(t)&=a_{0}e^{\pm\sqrt{\frac{2\Lambda}{N(N-1)}}t}.
\end{eqnarray}
If we choose the plus sign we have an $(N+1)$-dimensional de-Sitter
space and if we choose the minus sign we have an $(N+1)$-dimensional
anti de-Sitter space.

Notice that, since we have taken the separation constants $C_3$ and
$Q$ to be zero, the shape function~(\ref{b(r)}) is also the solution
for the static $(N+1)$-dimensional generalization of
zero-tidal-force wormhole models studied in Ref.~\cite{jamil}. The
analytical $(N+1)$-dimensional extensions for the energy density
$\rho(r)$ and pressures $p_r=\rho(r)^{\gamma}$ and $p_l(r)$ may be
directly obtained from Eqs.~(\ref{rho(r)})-(\ref{pl}) by making
$a(t)=const=1$.

Now we shall consider the conditions for having a wormhole
configuration (equally valid for evolving as well as for static
wormhole geometries). The condition $b(r_{0})=r_{0}$ allows us to
conclude that the integration constant $F$ may be written as
\begin{eqnarray}\label{F15}
F&=&-\frac{(\kappa \tilde{C}_{1}) ^{\frac{(\gamma-1)}{\gamma}}(N-2)^{1/\gamma}}{N(-\omega)^{1/\gamma}}r_{0}^{\frac{N(\gamma-1)}{\gamma}}+\nonumber\\
&&\left(\frac{(N-1)r_{0}}{2}\right)^{\frac{(N-2)(\gamma-1)}{\gamma}}.
\end{eqnarray}
From Eqs.~(\ref{b(r)}) and~(\ref{F15}) we obtain for the shape
function \small
\begin{eqnarray}\label{b(r)full}
b(r)=\frac{2r^{3-N}}{N-1}\Big[\frac{(\kappa \tilde{C}_{1})
^{\frac{(\gamma-1)}{\gamma}}(N-2)^{1/\gamma}}{N(-\omega)^{1/\gamma}}\Big(r^{\frac{N(\gamma-1)}{\gamma}}-r_{0}^{\frac{N(\gamma-1)}{\gamma}}\Big)\nonumber\\
+\left(\frac{(N-1)r_{0}^{(N-2)}}{2}\right)^{\frac{\gamma-1}{\gamma}}\Big]^{\frac{\gamma}{(\gamma-1)}}.\,\,\,
\end{eqnarray}
\normalsize Using the flare-out condition $b'(r=r_{0})<1$ we obtain
the following inequality:
\begin{eqnarray}\label{39}
\frac{2(3-N)}{N-1}+\frac{2(N-2)^{1/\gamma}(\kappa
\tilde{C}_{1})^{(\gamma-1)/\gamma}}{(N-1)(-\omega)^{1/\gamma}}r_{0}^{2(\gamma-1)/\gamma}
< 1.  \,\,\,\,\,\,
\end{eqnarray}
This is the $(N+1)$-dimensional extension of the flare-out condition
studied for polytropic static wormholes of the Ref.~\cite{jamil}.

If we consider standard phantom matter we must put $\gamma=1$. Thus
Eq.~(\ref{39}) implies that
\begin{eqnarray}
\frac{2(N-2)}{\omega}&>&7-3N.
\end{eqnarray}
Clearly, for $N=3$ we obtain that $\omega<-1$, i.e. the standard
constraint of the state parameter of a phantom matter component. In
$(3+1)$-dimensions, if $\gamma \neq 1$ then the state parameter
$\omega$ is allowed to take values $\omega > -1$, implying that we
have wormholes for phantom matter ($\omega<-1, C_1>0$), dark energy
($-1<\omega<-1/3, C_1>0$), matter with $-1/3< \omega< 0, C_1>0$, and
matter with $\omega > 0, C_1 < 0$.

It is interesting to note that Eqs.~(\ref{b(r)})-(\ref{pl}) imply
that if we require $\gamma<N/(N-1)$ then, in any dimension, the
shape function is asymptotically proportional to $r^{3}$, and the
energy density and pressures become time dependent only. Now, if
$\gamma>N/(N-1)$, the shape function (\ref{b(r)}) asymptotically
behaves as
\begin{eqnarray}\label{proporcional}
b(r)\propto r^{3-N}.
\end{eqnarray}
From this relation we conclude that for $N\geq3$ we have
asymptotically Minkowskian space-times. In this case, the space
slice $t=const$ of the metric~(\ref{2}) coincides with the space
slice of the $(N+1)$-dimensional extension of the Schwarzschild
black hole~\cite{BHN}.

On the other hand, if $F=0$, from Eqs.~(\ref{b(r)})-(\ref{pl}) we
obtain
\begin{eqnarray}
b(r)&=&\frac{2\kappa \tilde{C}_{1}}{N-1}\left(\frac{N-2}{N^{\gamma}(-\omega)}\right)^{\frac{1}{\gamma-1}}r^{3},\\
\rho(t)&=&\tilde{C}_{1}\left(\frac{N-2}{N(-\omega)}\right)^{\frac{1}{\gamma-1}}a^{-2}(t),\\
p_{r}(r)&=&-\tilde{C}_{1}\left(\frac{N-2}{N(-\omega)}\right)^{\frac{\gamma}{\gamma-1}}a^{-2}(t),\\
p_{l}(t)&=&-\tilde{C}_{1}\left(\frac{N-2}{N(-\omega)^{1/\gamma}}\right)^{\frac{\gamma}{\gamma-1}}a^{-2}(t).
\end{eqnarray}
where the scale factor is given by Eq.~(\ref{scalef}). In this case
the energy density and the pressures only depend on the time
coordinate, and the metric behaves like a FRW-metric with
$k=-1,0,1$.

Now, in order to shed some light on the properties of the discussed
solution~(\ref{b(r)})-(\ref{scalef}), we shall consider a particular
$(3+1)$-dimensional example by putting $\gamma=2$. As we can see
from Eq.~(\ref{b(r)full}), this value for the polytropic index
$\gamma$ ensures that the shape function is positive in any
dimension $N \neq 1$. Thus for $N=3$ we obtain \small
\begin{eqnarray}
b(r)&=&\left|\frac{\kappa
\tilde{C}_{1}}{9\omega}\right|r^{3}\Big[1+\left(\frac{r_{0}}{r}
\right)^{\frac{3}{2}}\left(3r_{0}^{-1}\sqrt{\frac{-\omega}{\kappa \tilde{C}_{1}}}-1\right)\Big]^{2},\label{b}\\
\rho(t,r)&=&\frac{
a_{0}\tilde{C}_{1}}{3|\omega|}\Big[1+\left(\frac{r_{0}}{r}\right)^{\frac{3}{2}}
\left(3r_{0}^{-1}\sqrt{\frac{-\omega}{\kappa \tilde{C}_{1}}}-1\right)\Big]\times\nonumber\\
&&e^{\pm\sqrt{\frac{\Lambda}{3}}t},\label{rhorho}\\
p_{r}(r,t)&=&-
a_{0}\left|\frac{\tilde{C}_{1}}{9\omega}\right|\Big[1+\left(\frac{r_{0}}{r}\right)^{\frac{3}{2}}
\left(3r_{0}^{-1}\sqrt{\frac{-\omega}{\kappa \tilde{C}_{1}}}-1\right)\Big]^{2}\times\nonumber\\
&&e^{\pm\sqrt{\frac{\Lambda}{3}}t},\\
p_{l}(t,r)&=&a_{0}\left|\frac{\tilde{C}_{1}}{18\omega}\right|\Big[1+\left(\frac{r_{0}}{r}\right)^{\frac{3}{2}}
\left(3r_{0}^{-1}\sqrt{\frac{-\omega}{\kappa \tilde{C}_{1}}}-1\right)\Big]\times\nonumber\\
&&\Big[-2+\left(\frac{r_{0}}{r}\right)^{\frac{3}{2}}\left(3r_{0}^{-1}\sqrt{\frac{-\omega}{\kappa \tilde{C}_{1}}}-1\right)\Big]\times\nonumber\\
&&e^{\pm\sqrt{\frac{\Lambda}{3}}t}.  \label{b5}
\end{eqnarray}
\normalsize

\begin{figure}[H]
    \centering
        \includegraphics[scale=0.2]{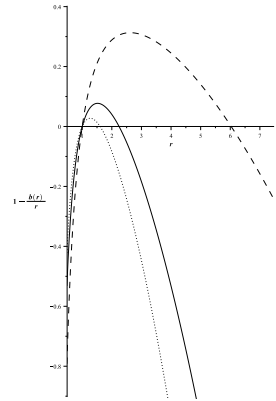}
    \caption{This figure shows the qualitative behavior of the metric component $g_{rr}^{-1}=1-\frac{b(r)}{r}$ with the shape function given by Eq.~(\ref{b}), for $\gamma=2$
    and different values of the state parameter $\omega$. For the shown curves we have used $r_{0}=1$ and $\kappa=1$. The dashed and
    solid lines represent the behavior of $g_{rr}^{-1}$ with $\tilde{C}_{1}=\frac{1}{5}$ for $\omega=-2$
    and $\omega=-\frac{1}{2}$ respectively. The dotted line represents
    the behavior of
    $g_{rr}^{-1}$ for $\omega=\frac{1}{3}$ and $\tilde{C}_{1}=-\frac{1}{5}$. The used parameter values satisfy the flare-out condition~(\ref{foc p}).
    The metric has the right signature $- +++$ for all $0<r_{0}<r<r_{b}$, allowing to exist wormhole configurations in this radial interval.}
    \label{figure-1}
    \end{figure}

\begin{figure}[H]
    \centering
        \includegraphics[scale=0.3]{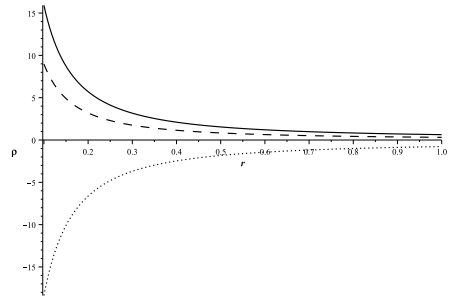}
    \caption{This figure shows the energy density $\rho(t_0,r)$, where $t_0=const$, given by Eq. (\ref{rhorho}), for $\gamma=2$ and different values of the state parameter $\omega$.
    For the shown curves we have chosen $r_{0}=1$, $\kappa=1$ and $a_{0}=1$. The dashed and solid lines represent the behavior of the energy density for
    $\omega=-2$ and $\omega=-\frac{1}{2}$ respectively, choosing $\tilde{C}_{1}=\frac{1}{5}$. The dotted line represents the behavior of the energy
    density for $\omega=\frac{1}{3}$ and $\tilde{C}_{1}=-\frac{1}{5}$. }
    \label{figura2}
    \end{figure}

From the above equations, we see that the relevant parameters of the
solution are the state parameter $\omega$, the the location of the
throat $r_{0}$ and the integration constant $\tilde{C}_{1}$. The
flare out condition~(\ref{39}) in this case becomes
\begin{eqnarray}\label{foc p}
\sqrt{\frac{\kappa \tilde{C}_{1}}{-\omega}}\, r_{0}<1,
\end{eqnarray}
with the extra requirement $\tilde{C}_1 \omega <0$. The energy
density $\rho(t,r)$ will be positive for
\begin{eqnarray}
r&<&r_{0}\left(1-\frac{3}{r_{0}}\sqrt{\frac{-\omega}{\kappa
\tilde{C}_{1}}}\right)^{2/3}, \
\textrm{and} \ \tilde{C}_{1}<0, \ \omega>0,\,\,\,\,\,\,\,\,\,\\
r&>&r_{0}\left(1-\frac{3}{r_{0}}\sqrt{\frac{-\omega}{\kappa
\tilde{C}_{1}}}\right)^{2/3}, \  \textrm{and} \ \tilde{C}_{1}>0,
\ \omega<0,\,\,\,\,\,\,\,\,\,  
\end{eqnarray}
and negative for
\begin{eqnarray}
r&>&r_{0}\left(1-\frac{3}{r_{0}}\sqrt{\frac{-\omega}{\kappa
\tilde{C}_{1}}}\right)^{2/3}, \
\textrm{and} \ \tilde{C}_{1}<0, \ \omega>0,\,\,\,\,\,\,\,\,\,\\
r&<&r_{0}\left(1-\frac{3}{r_{0}}\sqrt{\frac{-\omega}{\kappa
\tilde{C}_{1}}}\right)^{2/3}, \  \textrm{and} \ \tilde{C}_{1}>0, \
\omega<0. \,\,\,\,\,\,\,\,\,  \label{ed negative}
\end{eqnarray}
By using Eqs.~(\ref{b})-(\ref{ed negative}) qualitative plots for
$g^{-1}_{rr}$ and the energy density $\rho(t_0,r)$, where
$t_0=const$ are shown in Figs.~\ref{figure-1} and~\ref{figura2}. We
can see from Fig.~\ref{figure-1} that wormhole configurations are
allowed to exist only for $r_{0}<r<r_{b}$, where at $r=r_{b}$ is
located the mouth of the wormhole. This is because outside of this
radial interval the $g_{rr}$ metric component changes its sign, so
the metric holds the appropriate signature only at the radial
interval $r_{0}<r<r_{b}$. In the Fig.~\ref{figura2} we show that it
is possible to have wormholes with positive energy density for any
value $\omega<0$, while for $\omega>0$ the energy density is always
negative.

\section{Discussion}
In this paper we have obtained evolving $(N+1)$-dimensional wormhole
solutions supported by polytropic matter. The considered wormhole
models are described by a constant redshift function and generalizes
the standard flat FRW spacetime. The matter source is defined by the
``partially polytropic equation of state"~(\ref{pr}), where the
functions $\rho_{c}$ and $\rho_{w}$, depending on the time and
radial coordinates respectively, have been introduced. This
``partially polytropic equation of state" allows us to consider
generalizations of the polytropic wormholes discussed in
Ref.~\cite{jamil}. The generalization goes in two ways: first, we
obtain a static $(N+1)$-dimensional extension of zero-tidal-force
wormholes discussed by the authors of the mentioned
Ref.~\cite{jamil}, and secondly a dynamic extension of the same
static solution is obtained by introducing a scale factor with the
help of the metric~(\ref{15}). The considered field equations lead
to exponential expansion (contraction) of the scale factor due to
the presence of the cosmological constant.

We discuss a particular $(3+1)$-dimensional expanding solution with
the polytropic index $\gamma=2$. We show that it is possible to have
wormhole geometries with positive energy density for any value
$\omega<0$. For $\omega>0$ the energy density becomes always
negative.

Finally, notice that from expression~(\ref{b(r)}) we conclude that
if $F \neq 0$, $N \geq 3$ and $0 < \gamma \leq 1$, the solution is
asymptotically flat FRW space-time at spatial infinity, since by
requiring that $r \rightarrow \infty$ we obtain that $b(r)/r \approx
r^{2-N} \rightarrow 0$. By taking into account Eq.~(\ref{scalef}) we
conclude that the obtained asymptotic model corresponds to
N+1-extension of the de-sitter cosmology for $\Lambda>0$. The Hubble
rate is constant in any dimension. In standard 3+1-cosmology, the
de-Sitter model is used to describe the early universe during cosmic
inflation~\cite{Burgess}, and also the current accelerating
expansion in the framework of the $\Lambda$CDM model~\cite{LCDM}. By
neglecting ordinary matter, the dynamics of the universe is
dominated by the cosmological constant, dubbed dark energy, and the
expansion is exponential. The $\Lambda$CDM model has become the
standard model for modern cosmology, since is the simplest model
consistent with current observations.

In general, for $F \neq 0$, $N \geq 3$ and $0 < \gamma \leq 1$, the
obtained wormhole geometries far from the throat look like a flat
FRW Universe. If the wormhole throat is located outside of the
cosmological horizon of any observer, then he is not in causal
contact with the throat. Thus, an observer located too far from the
wormhole throat will see the Universe isotropic and homogeneous, and
in principle he will be unable to make a decision about whether he
lives in a space of constant curvature or in a space of a wormhole
spacetime~\cite{Cataldo15}.

\section{Acknowledgements}
This work was partially supported by CONICYT through Grant FONDECYT
N$^0$ 1080530 and by the Direcci\'on de Investigaci\'on de la
Universidad del Bio-B\'\i o through grants N$^0$ DIUBB 121007 2/R
and N$^0$ GI121407/VBC (MC).

\end{document}